\documentclass[aps,prl,twocolumn,showpacs,superscriptaddress]{revtex4}
\usepackage{dcolumn} \usepackage{bm} \usepackage{graphicx} \usepackage{amsmath}
\usepackage{latexsym} \usepackage{bbold} \usepackage{color} \usepackage{amsfonts}
\usepackage{amssymb} \usepackage{array} \usepackage{times} \usepackage{epsfig}
\usepackage{epstopdf} \usepackage{setspace} \usepackage{dsfont}

\def\I {\mathds{1}}

\newcommand\be{\begin{equation}}
\newcommand\ee{\end{equation}}
\newcommand\bea{\begin{eqnarray}}
\newcommand\eea{\end{eqnarray}}
\newcommand\ba{\begin{eqnarray*}}
\newcommand\ea{\end{eqnarray*}}
\newcommand\tr{\mbox{Tr}}

\newcommand\ket[1]{\left|#1\right\rangle}
\newcommand\bra[1]{\left\langle#1\right|}

\newcommand\ben{\begin{enumerate}}
\newcommand\een{\end{enumerate}}
\newcommand\bit{\begin{itemize}}
\newcommand\eit{\end{itemize}}

\newcommand{\cl}{{\hbox{\footnotesize cc}}}
\newcommand{\me}{{\hbox{\footnotesize me}}}
\newcommand{\dpw}{{\hbox{DP}}}
\begin{document}
\title{Discording power of quantum evolutions}
\author{Fernando Galve}
\affiliation{IFISC (UIB-CSIC), Instituto de F\'{\i}sica
Interdisciplinar y Sistemas Complejos, UIB Campus, E-07122 Palma
de Mallorca, Spain}
\author{Francesco Plastina}
\affiliation{Dipartimento di  Fisica, Universit\`a della Calabria,
I-87036 Arcavacata di Rende (CS), Italy}
\affiliation{INFN - Gruppo collegato di Cosenza}
\author{Matteo G. A. Paris}
\affiliation{Dipartimento di Fisica dell'Universit\`a degli Studi di
Milano, I-20133 Milano, Italy.}
\affiliation{CNISM, UdR Milano Statale, I-20133 Milano, Italy}
\author{Roberta Zambrini}
\affiliation{IFISC (UIB-CSIC), Instituto de F\'{\i}sica
Interdisciplinar y Sistemas Complejos, UIB Campus, E-07122 Palma
de Mallorca, Spain}
\date{\today}
\begin{abstract}
We introduce the discording power of a unitary transformation,
which assesses its capability to produce quantum discord, and
analyze in detail the generation of discord by relevant classes
of two-qubit gates. Our measure is based on the Cartan
decomposition of two-qubit unitaries and on evaluating the maximum
discord achievable by a unitary upon acting on classical-classical
states at fixed purity. We found that there exist gates which are
perfect discorders for any value of purity, and that they belong
to a class of operators that includes the $\sqrt{\mbox{SWAP}}$.
Other gates, even those universal for quantum computation, do not
posses the same property: the $\mbox{CNOT}$, for example, is a
perfect discorder only for states with low or unit purity, but not
for intermediate values. The discording power of a two-qubit
unitary also provides a generalization of the corresponding
measure defined for entanglement to any value of the purity.
\end{abstract}
\maketitle
The primary aim of the science of quantum information is the
exploitation of the quantum structure of nature for information
processing and communication tasks. Among the quantum features of
a physical system, entanglement is usually considered the
prominent resource,  providing speed-up in various quantum
information and communication tasks \cite{horormp}.  In the realm
of mixed-state quantum-information, however, instances are known
where quantum advantages are obtained in the presence of little or
no entanglement. In fact, quantum discord \cite{discord,henderson}
has been proposed as the source behind this enhancement
\cite{lanyon2008}, and some indications in this direction have
been given \cite{eastin,fanchini,shaji11,discreview}. The notion
of nonclassicality springing from information theory has been also
discussed in comparison with that coming from phase-space
constraints \cite{fer12}.
\par
Although introduced in the context of environment induced
decoherence, quantum discord has been then related to the
performance of quantum and classical Maxwell's demons \cite{demon}
and to the thermodynamic efficiency of a photo-Carnot engine
\cite{dille}, as well as the total entanglement consumption
\cite{jquindici}, and the minimum possible increase of quantum
communication cost \cite{jquattordici} in state merging protocols.
Furthermore, its propagation properties have been studied in
\cite{borrelli12} in connection with micro-causality and in
\cite{campbell11} for quantum spin channels. \par Quantum discord
can be activated into distillable entanglement
\cite{jdiciotto,oltre18}, and has been shown to be a resource in
quantum state discrimination \cite{reso} and quantum locking
\cite{lock}. It has been also shown \cite{celeri,madhokarx} that
discord quantifies the minimum damage made by a decoherence
process to the performance of many quantum information processing
protocols, including teleportation, distillation and dense coding.
\par
In continuous variable systems, Gaussian quantum discord
\cite{gio10,ade10} has been experimentally measured
\cite{expD1,expD2,expD3} and represents  a measure of the
advantage provided by coherent quantum systems over their
classical counterpart. It has been also suggested \cite{semprevla}
that the geometric  quantum discord \cite{geo} is the optimal
resource for remote quantum state preparation, though the
interpretation of geometric discord as a measure of correlations
has been questioned \cite{MP}.
\par
This body of recent knowledge represents a strong motivation
to understand in quantitative terms how well quantum discord
may be produced by a given operation. To this aim we focus on
two-qubit systems and introduce the discording power
of (non-local) unitary gates, a quantity which allows us to investigate
in detail the controlled production of symmetric discord.
In particular, the main question we want
to answer is the following: which is the maximum discord
that a gate may produce acting on classical-classical
states \cite{PHH}, i.e. states with zero discord ?
\par
In order to answer this question, we define below the discording
power $\dpw_{\!\mu}[U]$ of a given two-qubit gate $U$, as the
maximum amount of (symmetric) discord produced by $U$ when acting
on the set of classical-classical states $\rho_{\cl}^\mu$, i.e.
the set of states with zero discord and fixed purity
$\mu=\tr\left[\left( \rho_{\cl}^{\mu}\right)^2\right]$. The set of
two-qubit unitaries is thus naturally split into equivalence
classes, each individuated by the discording power, represented by
curves in the purity-discord plane. Notice that the discording
power provides a generalization of the so-called entangling power
of gate $U$, \cite{kraus,gene}, which is obtained as
$\dpw_{\!\mu=1}$ since, for pure states, quantum discord coincides
with entanglement.
\par
Thanks to the fact that Maximally Discordant Mixed States (MDMS)
for a given purity have been identified in \cite{alqasimi} (see
also \cite{roberta}), we are able  to find analytically the class
of gates that are perfect, i.e. the {\em best-discorder}, for a
given purity. These are defined as those unitaries that produce a
MDMS by acting upon some $\rho_{\cl}^\mu$. In this way, one also
sees that the notion of best-discorder is the generalization to
any value of the purity of the perfect entangler discussed in
\cite{bestent}. As a first result, we found that there exist gates
which are perfect discorders for every purity $\mu$. They pertain
to a class of operators that includes the $\sqrt{\mbox{SWAP}}$.
This is by no means a trivial property, since many gates do not
hold it, not even all of the two-qubit gates that are universal
for quantum computation. For example, the $\mbox{CNOT}$ gate (as
well as the other unitaries to which it is equivalent in the sense
specified below) is a perfect discorder for very high or low
purity, but not for intermediate values. More specifically, CNOT
works as a perfect discorder for rank-$4$ and rank-$3$ states, and
it is a perfect entangler as well (that is, a perfect discorder)
for rank-$1$ states. However, it fails to achieve the rank-$2$
MDMS, that are, instead, obtained after the action of
$\sqrt{\mbox{SWAP}}$ on a suitable rank-$2$ $\rho_{\cl}$
\cite{rank}.
\par
To proceed with the formal introduction of our figure of merit,
we first recall that the discord can be
understood as the difference between the mutual information and
the classical correlations, \cite{henderson}. For a generic
bi-partite state $\rho_{AB}$, the mutual information is given by
${I}[\rho_{AB}] = S[\rho_A] + S[\rho_B] - S[\rho_{AB}]$, where
$\rho_A$ and $\rho_B$ are the local reduced density operators and
$S[\rho]$ the von-Neumann entropy of the state $\rho$. As
for the classical correlation, we suppose that a POVM-measurement
${\cal M}_B$ with elements $\{ M_k \}$ (with $M_k \geq 0$ and
$\sum_k M_k = \I$) is performed on qubit $A$. It realizes for $B$
the post-measurement ensemble ${\cal E}_B= \{ p_k, \rho_{B}^{k}
\}$ where $p_k = \tr \{ \rho_{AB}(M_k\otimes \I) \}$ is the
probability for the $k$-th outcome, while the $k$-th
post-measurement state of $B$ is $\rho_{B}^{k} = \tr_A
\{\rho_{AB}(M_k \otimes \I)\}/p_k$. The amount of information
acquired about qubit $B$, optimized over all possible POVM, gives
the classical correlations $C^{AB}[\rho_{AB}] =
S[\rho_B] - \min_{{\cal M}_B} \sum_k p_k S[\rho_{B}^{k}]$.
The (one-way) quantum discord is then \be \delta^{AB}[\rho_{AB}] :=
{I}[\rho_{AB}] - C^{AB}[\rho_{AB}] \, .\ee An analogous procedures leads to the
definition of discord with measurement on $B$, $\delta^{BA}$.
\par
Since the conditional entropy is concave over the convex set of
POVMs, the minimum is attained on the extreme points of the set,
having rank-$1$ \cite{rankuno,j34}. Discord is typically evaluated
in a simplified form, where only orthogonal measurements are
considered, rather than the more general POVM. For two-qubit
states, this is enough to achieve the minimum for rank-$2$ states;
while, for rank-$3$ and -$4$ states orthogonal measurements give a
pretty tight upper bound, as shown in Ref. \cite{almost}. Given
the numerical evidence provided there, the improvement in doing
full minimization is on average at the level of $10^{-6}$.
Therefore, in the following, we will restrict the evaluation of
discord to projective measurements.
\par
We define the {\it discording power} of a gate ${U}$ as the
maximum symmetrized quantum discord that can be achieved by such
gate from {\it any} classical-classical state of a given purity
$\mu$ ($1/4 \leq \mu \leq 1$):
\begin{equation}
\dpw_{\!\mu} [{U}]\equiv \max_{ \rho_{\cl}^{\mu}} \: \delta[U
\rho_{\cl}^{\mu} U^{\dag}] \, .
\end{equation} Here $\delta$ is the symmetrized discord, \begin{equation}
\delta[U \rho_{\cl}^{\mu} U^{\dag}] = \frac{1}{2} \left(
\delta^{AB}[U \rho_{\cl}^{\mu} U^{\dag}]+ \delta^{BA}[U \rho_{\cl}^{\mu}
U^{\dag}]\right)  \notag \, , \end{equation} while a (concordant)
classical-classical state $\rho_{\cl}^{\mu}$ corresponds to a
convex combination of orthogonal projectors:
\begin{equation}
 \rho_{\cl}^{\mu}= \sum_{r,s} p_{r,s}
\ket{\alpha_r}\bra{\alpha_r}\otimes\ket{\beta_s}\bra{\beta_s} ,
\mbox{with} \, \sum_{r,s} p_{r,s}^2 = \mu \label{rhoclas}
\end{equation}
where $\ket{\alpha_r}$, $\ket{\beta_s}$ are elements of orthonormal
basis for the two qubits, and the $\{p_{r,s}\}$ are probability
distributions.
\par
Any two-qubit unitary may be written in Cartan form \cite{kraus}
${U}=\left({L}_1\otimes {L}_2\right) \, U_c(\vec{\theta}) \,
\left({L}_3\otimes {L}_4\right)
$
with
\begin{equation}
U_c(\vec{\theta}) =
\exp\left(-i\sum_{j=x,y,z}\theta_j{\sigma}_j\otimes{\sigma}_j\right)
\, ,
\end{equation}
where $L_i$ are  local unitaries, and where (due to symmetry
reasons) the independent values of the $\theta_j$ are constrained by
$0 \leq \theta_z \leq \theta_y \leq \theta_x \leq \pi/4$ \cite{notaC}.
\par
At first, we notice that $L_1$ and $L_2$ are not affecting the
value of $\dpw_{\!\mu}[U]$  since the discord is left unchanged by
local unitary operations. The local unitaries $L_3$ and $L_4$ are
irrelevant as well because $\dpw_{\!\mu}[U]$ is the result of a
maximization over classical-classical states at fixed purity. This
involves a scan over the distributions $p_{r,s}$  and the
orthonormal qubit basis in Eq. (\ref{rhoclas}), and since the
latter are just rotations of the logical basis $\{\ket{0},
\ket{1}\}$, we can remove $L_{3,4}$ by combining them with the
rotations needed to check all local basis. Overall, we have that
all of the gates with a given Cartan kernel $U_c$ have the same
discording power, which, itself, can be intended as a function of
the three parameters $\theta_j$.
\par
\begin{figure}[h]
\includegraphics[width=0.98\columnwidth]{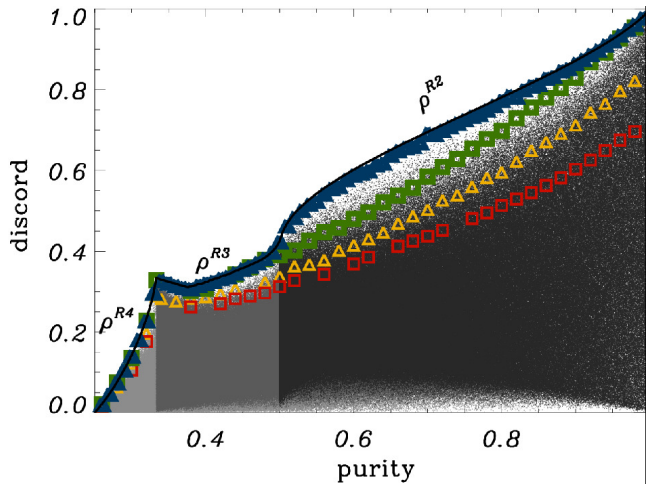}
\caption{(Color online) Symmetric quantum discord ($\delta$)
versus purity ($\mu$) for two-qubit unitary kernels.  The
maximum discord for a given purity (black continuous line)
is obtained for $\rho^{R2},\rho^{R3},\rho^{R4}$ (see main text).
The (numerically evaluated) $\dpw_{\!\mu}$ for
different gates is represented with symbols: from the top,
$U_c(\pi/8,\pi/8,0)$ (bold triangles), $U_c(\pi/4,0,0)$ (bold
squares), $U_c(\pi/4,\pi/4,0)$ (triangles), $U_c(0.15\pi,0,0)$
(squares). The numerical maximization for each gate and for any value of
$\mu$ is performed by considering {$\sim 8\cdot 10^6$ classical states
, with local rotation angles discretizing $\ket{\alpha_r}$
and $\ket{\beta_s}$ by steps of $0.1\pi$ and
 $\sim 5 \cdot 10^2$ different values for $p_{r,s}$}. The
agreement with the analytic result can be appreciated by
considering the two overlapping upper curves, corresponding to the
border states and $\dpw_{\!\mu}[U_c(\pi/8,\pi/8,0)]$. This serves the purpose
of a consistency check. On the
background, layers of $10^8$ random matrices of rank-$2$ (dark
points), -$3$ (intermediate grey) and -$4$ (lighter grey) are
superimposed. \label{fig1}}
\end{figure}
\begin{figure}[h]
\includegraphics[width=0.98\columnwidth]{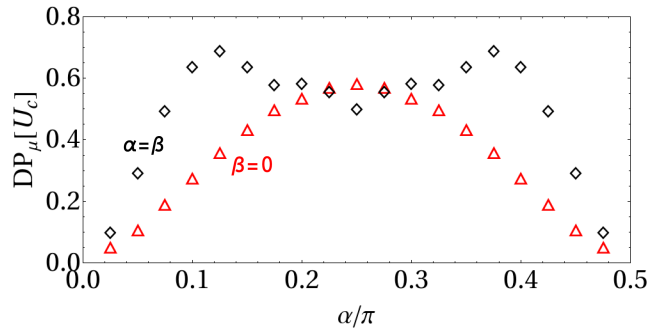}
\caption{Discording power of two families of gates: $\dpw_{\!\mu}
[U_c(\alpha,\beta=\alpha,0)]$ (diamonds symbols) and $DP_\mu[U_c(
\alpha,0,0)]$ (triangle symbols) for a fixed purity $\mu =0.7$ as
a function of the angle parameterizing the gates. \label{fig2}}
\end{figure}

Every two-qubit gate can be associated to a curve in the
discord-purity plane, given by $\dpw_{\!\mu}[U_c]$. Examples may
be seen in Fig. (\ref{fig1}), where we report the symmetrized
discord $\delta$ as function of purity, and in turn the discording
power, for several Cartan kernels of the form $U_c(\alpha, 0,0)$
and $U_c(\alpha, \alpha,0)$ (chosen here because of their
relevance for the discussion below). The discording power of these
kernels is also plotted in Fig. (\ref{fig2}) for a fixed value of
the purity, $\mu=0.7$ as a function of the angle $\alpha$. We can
see from these two plots that the gates $U_c(\pi/4,0,0)$ and
$U_c(\pi/8,\pi/8,0)$ have better performances in generating
discord. The first of them is the kernel of the $\mbox{CNOT}$
gate, and, from Fig. (\ref{fig1}), it can be appreciated to be a
perfect discorder for rank-$4$ and rank-$3$ states. The second one
(in fact, the whole class $U_c(\pi/8,\pi/8, \gamma)$, $\forall
\gamma$) is a best discorder too; but, this time, for the whole
range of purities and all ranks. In particular, the operator
corresponding to $\vec{\theta} = (\pi/8,\pi/8,\pi/8)$ gives the
$\sqrt{\mbox{SWAP}}$ up to an irrelevant phase.
\par
In order to discuss in more detail the properties of some unitary
to be a perfect discorder, we need to recall the expressions for
the MDMS, separately for each rank. To this end, in Fig.
(\ref{fig1}), the region of physically admissible states in the
$\delta$-$\mu$ plane is indicated by the background area. The
boundary of this region has been identified in Ref.
\cite{alqasimi} and is given by states, indicated as $\rho^{Rn}$,
with $n=2,3,4$, that  are either symmetric under the exchange of
$A$ and $B$ or related to their symmetric counterpart by a local
rotation. This means that both $\delta^{AB}$ and $\delta^{BA}$ (as
well as the symmetric discord $\delta$ that we have chosen to
employ) are maximized on this border.
\par
The boundary is quite composed: For low purities, only rank-$4$
states are present and the maximum discord for a given purity is
obtained by Werner states, $\rho^{R4}\equiv
\rho(w)=\frac{1-w}{4}\I+w\ket{\Psi_{\me}}\bra{\Psi_{\me}}$. Here,
$\ket{\Psi_{\me}}$ can be any maximally entangled state, while the
Werner parameter, that in principle lays in the range $[-1/3,1]$,
is here confined to negative values $w=-\sqrt{(4 \mu-1)/3}$, with
the purity restricted to $1/4\leq \mu \leq 1/3$. Interestingly,
for these values of the purity, the MDMS are separable states,
being this lack of entanglement a feature also found when
maximizing discord for a given classical correlation
\cite{roberta}. On the other hand, rank-$3$ states maximizing
discord for a given purity are entangled. They are obtained within
the family $\rho^{R3}\equiv\rho(a,b,\varphi)$ \cite{alqasimi},
where
$$\rho(a,b,\varphi) = \frac{1}{2}
\begin{pmatrix} 1-a+b & 0 & 0 & 0 \\ 0 & a & a e^{-i \varphi} & 0 \\
0 & a e^{i \varphi} & a & 0 \\
0 & 0 & 0 & 1-a-b \end{pmatrix} \, ,$$ for a proper choice of $a$
and $b$, with $a \in [0,1]$ and $|b| \leq 1-a$ (notice that
neither the discord nor the purity depend on  $\varphi$, as it can
be canceled by a rotation of qubit $A$ around the $z$ axis; we
inserted this phase here for future reference). Finally, rank-$2$
MDMS are obtained from $\rho^{R3}$ by taking $b=1-a$ and $1/2 \leq
a \leq 1$, i.e. $\rho^{R2} = \rho(a,1-a,\varphi)$. As noted in
\cite{francica}, these states are quite peculiar as their discord
is equal to their concurrence, $\delta(\rho^{R2}) = {\cal
C}(\rho^{R2}) = a$.
Both the Werner states and the rank-$3$ states $\rho^{R3}$ can be
obtained from a classical state under the action of both the
$\mbox{CNOT}$ and the $\sqrt{\mbox{SWAP}}$. To show this
explicitly, let us start by considering the classical-classical
state with the same eigenvalues as the Werner one: $\rho_{cl}^{R4}
= \mbox{diag} \{(1-w)/4,(1-w)/4,(1-w)/4,(1+3w)/4 \}$. It is, then
easy to see that
$$U_c(\pi/4,0,0) \, \rho_{cl}^{R4} \, U_c^{\dag}(\pi/4,0,0) =
\frac{1-w}{4} \I + w \ket{\Phi}\bra{\Phi} \, ,$$ which is a Werner
state since $\ket{\Phi} = (\ket{00} + i \ket{11})/\sqrt 2$ is
maximally entangled. This explains the first part of the plot in
Fig. \ref{fig1} in which the bold squares (representing
$U_c(\pi/4,0,0)$) fall on the MDMS border. The same is true for
any operator with a kernel of the form $U_c(\pi/8,\pi/8,\gamma)$
as, by considering a rotated classical state $\rho_{cl}^{R4'} =
(\sigma^x \otimes \I) \rho_{cl}^{R4} (\sigma^x \otimes \I)$, one
has that $U_c(\pi/8,\pi/8,\gamma) \, \rho_{cl}^{R4'} \,
U_c^{\dag}(\pi/8,\pi/8,\gamma)$ gives a Werner state with
$\ket{\Psi_{me}}= (\ket{01} + i \ket{10})/\sqrt 2$. It can be seen
by looking at bold triangles in Fig. (\ref{fig1}), that this
operator is a perfect discorder for any rank. In fact, one has
$$U_c(\pi/8,\pi/8,\gamma) \, \rho_{cl}^{R3} \,
U_c^{\dag}(\pi/8,\pi/8,\gamma)= \rho(a,b,\varphi=\pi/2) \, ,$$
where $\rho_{cl}^{R3} = \mbox{diag} \{(1-a+b)/2,0,a,(1-a-b)/2\}$.
This shows that, for any value of $\gamma$, the gate
$U_c(\pi/8,\pi/8,\gamma)$ is able to reach the MDMS border of both
rank-$3$ and rank-$2$, thus showing that this is a perfect
discorder for every value of the purity.
\par
On the other hand, the $\mbox{CNOT}$ gate (and all of the
operators with the same Cartan kernel), although able to reach the
rank-$3$ MDMS, fails for rank-$2$ states. This impossibility can
be shown analytically by considering its reverse action on
$\rho^{R2}$: even by allowing local rotations,
$U_c(\pi/4,0,0)^{\dag}$ applied on $\rho^{R2}$ gives an entangled
(and, thus, non-classical) state. This clearly implies that the
action of $U_c(\pi/4,0,0)$ is not enough to obtain the rank-$2$
MDMS acting on classical states. We also notice that one of the
main differences between discord and entanglement is that the
former can be increased by local (nonunitary) operations. On the
other hand, MDMS of rank-2 and -3 are entangled and thus cannot be
created by local operations on classical-classical states. As a
consequence, full discording power can only be achieved by
nonlocal, two-qubit, gates. The case of rank-4 states remains an
open question though.
\par
In summary, we have introduced the discording power of a unitary
transformation, which measures its capability to produce quantum
discord, and have analyzed in detail the generation of symmetric
discord by relevant classes of two-qubit gates.  Our measure is
based on the Cartan decomposition of two-qubit unitaries and on
evaluating the maximum discord achievable from classical-classical
states at fixed purity.  We have identified the gates that are
able to generate MDMS, and discussed their performance as a
function of the purity.  We found that there exist gates which are
perfect discorders for any value of purity, and that they belong
to a class of operators that includes the $\sqrt{\mbox{SWAP}}$.
Even gates that are universal for quantum computation do not share
this property, e.g. the $\mbox{CNOT}$ is a perfect discorder for
rank-$4$ and rank-$3$ states, but it fails to achieve the rank-$2$
MDMS.
\par
The discording power of a two-qubit unitary provides a
generalization of the corresponding measure defined for
entanglement on pure states. Our results represent the first
attempt to evaluate the correlating power of gates for mixed
states, a topic which is largely unexplored even for entanglement
\cite{bat03}, and completely absent for other measures not related
to the separability paradigm. We also notice that similarly to
what is known for entanglement quantifiers, geometric discord and
quantum discord yield different states ordering in the mixed
setting \cite{yeo10,okrasa12}, thus probably inducing inequivalent
discording powers, let alone other less related nonclassicality
measures, \cite{discreview}. Our work thus also motivates the
search of an entangling power measure starting from mixed states,
in order to compare entanglers and discorders at a general value
of purity.
\par
The use of quantum
discord as a resource for quantum technology is still a debated
topic, and a definitive answer  may only come from experiments
involving carefully tailored states and operations. Our results go
in that direction providing a precise characterization of
two-qubit unitaries in terms of their ability in generating
quantum discord.
\par
This work has been supported by the visiting professorship program of
UIB, the CSIC post-doctoral JAE program, the MIUR project FIRB
LiCHIS-RBFR10YQ3H, the MICINN project FIS2007-60327 and the MINECO
project FIS2011-23526.


\begin {thebibliography}{99}
\bibitem{notaC}{It
was shown in \cite{kraus} that, up to local rotations, the Cartan
kernel $U_c$ is periodic with period $\pi/2$, and is
reflection symmetric around $\pi/4$. This, together with the arbitrariness
of $x,y,z$ labelling, yields the reduced range for $\theta_j$ that we have
chosen.}
\bibitem{rank}  Rank-$n$ states are states $\rho$ with a
rank-$n$ density matrix. The number $n$ is thus the dimension of
the subspace spanned by the eigenvectors of $\rho$. Pure states
are rank-1 states. Mixed states of two qubits may be rank-2,
rank-3 or rank-4 states. Purity of rank-$n$ states lies in the
range $[1/n,1]$.
\end{thebibliography}
\end{document}